\documentclass{article}
\usepackage[latin1]{inputenc}
\usepackage[english]{babel}
\usepackage{amsmath}
\usepackage{amsfonts}
\usepackage{amssymb}
\usepackage{graphicx}
\usepackage{authblk}
\usepackage{xcolor}
\usepackage{ulem}
\usepackage{multirow}
\usepackage{verbatim} 

\usepackage[T1]{fontenc}
\usepackage[font=small,labelfont=bf,tableposition=top]{caption}

\DeclareCaptionLabelFormat{andtable}{#1~#2  \&  \tablename~\thetable}

\title{MicroVelocity: rethinking the Velocity of Money for digital currencies}
\author[1,2,3]{Carlo Campajola}
\author[4]{Marco D'Errico\footnote{The views expressed in this paper are those of the authors and do not
necessarily reflect those of the ECB, the ESRB or its member institutions.}}
\author[2,3]{Claudio J. Tessone}

\affil[1]{DLT Science Foundation, London, United Kingdom}
\affil[2]{University of Zurich, Blockchain \& Distributed Ledger Technologies, Department of Informatics, Z\"urich, Switzerland}
\affil[3]{UZH Blockchain Center, Z\"urich, Switzerland}
\affil[4]{European Systemic Risk Board Secretariat, European Central Bank, Frankfurt, Germany}
\date{February 1, 2022}

\begin{document}

\maketitle

\begin{abstract}
    We propose a novel framework to analyse the velocity of money in terms of the contribution (MicroVelocity) of each individual agent, and to uncover the distributional determinants of aggregate velocity. Leveraging on complete publicly available transactions data stored in blockchains from four cryptocurrencies, we empirically find that MicroVelocity i) is very heterogeneously distributed and ii) strongly correlates with agents' wealth. We further document the emergence of high-velocity intermediaries, thereby challenging the idea that these systems are fully decentralised. Further, our framework and results provide policy insights for the development and analysis of digital currencies.
    
\end{abstract}

{\footnotesize \textbf{Keywords}: Velocity of Money, Cryptocurrency, Blockchain, Heterogeneous agents

 }

\section*{Introduction}

One of the earliest and most widespread concepts across societies is that of money. As humans we are inherently brought to organise ourselves into communities and exchange the goods and services we produce; this leads to the necessity to invent a medium of exchange that can be store of value - to be able to defer purchases to the future - and unit of account - to properly quantify the value of goods and services \cite{mishkin2004economics}. Any such artifact, which may take the form of anything between a large stone in the middle of the village \cite{fitzpatrick2020banking}, cigarettes \cite{burdett2001cigarette}, cash in the form of coins and banknotes or an encrypted sequence of bits \cite{nakamoto2008peer}, is a form of money, as long as there is more than one person willing to accept it as such.

Over the years, economic theory has kept interrogating itself on how to efficiently manage this peculiar asset: who is entitled to create it, what is its cost and, most importantly, the role of money on societal and economic development. These and other questions led to the development of monetary theory, which interrogates on how the quantity of money in a system influences economic activity and macroeconomic quantities.

One of the main research areas in monetary theory is the quantity theory of money, which revolves around Fisher's equation of exchange \cite{fisher2006purchasing}

\begin{equation}\label{eq:fisher}
    MV = PQ
\end{equation}

stating the proportionality between the monetary mass $M$ and the level of prices $P$ through the economic output $Q$ and $V$, the \textit{velocity of money}. The latter is defined as the average frequency with which a given unit of money is used to purchase domestically-produced goods and services within a given time period: this makes Fisher's equation an accounting identity which holds true by construction as both sides amount to the total money flow, which we will call $F$. The velocity $V$ is often seen as an indicator of money demand \cite{friedman1959demand}: if money demand is low - i.e. expected yield from alternative assets is high - economic agents will convert their money into less liquid investments as soon as possible, thus leading to increased velocity, whereas in situations of high demand for money the velocity will decrease. 

Money velocity is not directly measurable since it would be impossible to track every unit of currency circulating in the economy. Therefore, the equation of exchange provides a way to estimate velocity at a given point in time. The equation however holds true \textit{in aggregate}, thus providing limited insights on the \textit{distributional} aspects, in particular for what concerns a more granular measurement of velocity and how different economic agents (or groups of agents) contribute to the aggregate. In fact, agents are heterogeneous, have different spending habits, different levels of wealth, income and demand for money and different roles in its movement in the economy; but since their different contributions to aggregate velocity are not measurable, policy analysis and interventions cannot be targeted at a finer level.

However, these limitations to the measurement of velocity might be not so strict anymore: over the last few years the increasing diffusion of digital payments - via electronic cards, e-banking or smartphone apps, to name some alternatives - has led to a surge in transactions data availability, meaning that it is now within reach to actually measure the velocity of money from micro-level data, at least for digital forms of money. A big role in this transition towards digital cash will be played by central banks, who are now devoting significant resources towards the development of Central Bank Digital Currencies (CBDCs) \cite{engert2017central, bis2020report, ecb2020report} to provide citizens with access to central bank money in digital form. 

A strong push towards digital payments has been prompted by the rise over recent years of cryptocurrencies and blockchain-based tokens. These have emerged as a family of unconventional financial assets and some of them, with Bitcoin being the first and most prominent example, claimed they could be the future of money and would take over the role of fiat currencies. While this has not happened, and outside of crypto-maximalist communities it is now commonplace \textit{not} to consider Bitcoin and its many offsprings as currencies, the publicly available records of blockchains constitute a unique opportunity to explore the space of microeconomically grounded metrics for macroeconomic quantities like the velocity of money.

In particular in this paper we focus on exploiting the accounting protocol adopted by several cryptocurrencies, the \textit{Unspent Transaction Output} (UTXO) protocol, to actually track specific units of currency as they move from one agent to the other. This protocol imposes that whenever funds have to be spent they are included as \textit{inputs} of a transaction, and they have explicit reference to the unspent \textit{output} of a previously validated transaction. This provides verification that the issuer of the new transaction has the right to spend those funds, as the output can only be accessed by means of a cryptographic private key owned by the receiver. A by-product of this mechanism is that every input to a new transaction is explicitly linked to the output of a previous one, and thus the time that those units of money stayed idle is known.

As we explain in the following, this level of detail allows to factor the total velocity $V$ into agent-specific contributions, $V_i$, which we call \textit{MicroVelocity} and depend on agents' behavioral usage of their wealth. This description, based on the seminal work of   \cite{wang2003circulation}, opens to several questions: how is MicroVelocity distributed cross-sectionally? How does it evolve over time? How does an agent's MicroVelocity correlate with their wealth? How does it correlate with the MicroVelocity of their partners in transactions?

In the context of cryptocurrencies, the velocity of money has been used to propose valuation models \cite{athey2016bitcoin, bolt2020value}, by an argument stating that if the economy adopting cryptocurrency as its medium of exchange grows, so does the velocity due to the available supply being limited by protocol, thus an increase in velocity should also warrant an increase in pricing. Several estimators have been proposed for the velocity of cryptocurrencies, like Coin Days Destroyed \cite{ciaian2016digital} or Average Dormancy \cite{smith2018bitcoin}. In particular Pernice et al. \cite{pernice2020cryptocurrencies} propose an accurate method to estimate velocity on aggregate, which takes advantage of the UTXO protocol in similar ways to our definition of MicroVelocity.

We argue that MicroVelocity allows more detailed considerations on the state of the economy using cryptocurrencies. Specifically, statistical properties of its distribution across agents provide insight into the global level of specialization and on centralization in the flow of money: if the distribution is highly heterogeneous, agents with high MicroVelocity are likely to be intermediaries and those with low $V_i$ their clients, thus going against the common narrative that cryptocurrencies are ``decentralized economies" \cite{makarov2020trading}. Paths towards the centralization of these systems are currently attracting growing attention, with Aramonte et al. \cite{aramonte2021defi} explicitly talking of ``decentralization illusion".

Regardless, our framework is not necessarily limited to the context of blockchain-based tokens. The formulation of MicroVelocity general in nature and can be adapted to find application to any payment system that keeps a digital record of transactions, such as bank transfers, credit cards payments or, in the future, CBDCs. Such an instrument will then be a useful addition to economists' toolboxes when investigating effects of policy, and possibly even opening new policy channels that account for the heterogeneity of economic agents in their usage of money \cite{schnabel2020unequal, ahnert2020understanding}.

\paragraph*{MicroVelocity.}
We build on Wang and coauthors' theoretical framework \cite{wang2003circulation}, which shows that the macroeconomic velocity can be derived from a microeconomic description: here we review and expand their argument, introducing our definition of MicroVelocity.

Consider the time between two transactions involving the same unit of currency, the \textit{holding time} $\tau$, and assume that $\tau$ is a random variable with probability distribution $p(\tau)$. A condition $p(0)=0$ has to be enforced since a holding time of $0$ is meaningless for the description. It follows that, calling $M$ the total number of coins in circulation, $Mp(\tau)d\tau$ is the number of coins with a holding time in the interval $[\tau, \tau + d\tau)$.

Assuming stationarity of $p(\tau)$, the money flow $F(\tau)$ generated by coins with holding time $\tau$ is

\begin{equation}
    F(\tau) = M \frac{1}{\tau} p(\tau) d\tau
\end{equation}

as $1/\tau$ is the frequency with which each coin moves. Since the total economic flow $F$ is the integral of this quantity over all holding times, we find

\begin{equation}\label{eq:flow}
    F = MV = PQ = M \int_0^{\infty} \frac{1}{\tau} p(\tau) d\tau
\end{equation}

and it thus follows that the velocity reads

\begin{equation}\label{velocitywang}
    V = \int_0^{+\infty} \frac{1}{\tau} p(\tau) d\tau
\end{equation}

It has to be noticed that units of currency are not distinguishable and there is no reason to think that a particular coin has an inherently different holding time than any other one. Indeed the authors of \cite{wang2003circulation} justify the existence of the holding times distribution by arguing that holding times depend on money ownership, \textit{i.e.} the spending habits of different economic agents produce the heterogeneity in $\tau$. While coins are indistinguishable, economic agents are not and each has their own spending, saving and speculative attitude. 

In the original paper the authors only consider the aggregate effect of this heterogeneity, defining $p(\tau)$ but neglecting to consider the individual effects of agents, as a simulation study accounting for it would have introduced too many arbitrary parameters in the description. Our goal is thus to overcome this limitation, as having access to agent-level data allows for a more fine-grained representation. 

Let us then define $p_i(\tau)$ as the probability distribution of holding times of agent $i$, with $i \in \lbrace 1,2,\dots N \rbrace$. This means that the total flow of Eq. \ref{eq:flow} is factored in a sum of individual contributions, and then Eq. \ref{velocitywang} transforms into

\begin{equation}\label{eq:microvel}
    V = \sum_{i=1}^N V_i = \sum_{i=1}^N \frac{M_i}{M} \int_0^\infty \frac{1}{\tau} p_i(\tau) d\tau
\end{equation}

where $M_i$ is the amount of money held by agent $i$, meaning that the total velocity is a \textit{wealth-weighted} average of individual contributions, the MicroVelocity $V_i$. This equation makes it clear that, in a population with homogeneous spending preferences such that $p_i(\tau) \approx p(\tau)$ $\forall i$, contributions to money velocity are fully determined by wealth. However real-world economies involve heterogeneous participants, both on the macro-scale (households, companies, business sectors) and on the micro-scale (preferences, consumption functions): our framework fully accounts for these factors, returning a detailed picture of the impact they have on the velocity.

\paragraph*{Blockchains and the UTXO protocol.}
To estimate MicroVelocity from real-world data one needs fine-grained transaction records, so that the holding times $\tau$ can be measured and its statistical properties estimated. The advent of blockchains and cryptocurrencies offers a unique occasion to study economic activity on a publicly available transaction ledger. A blockchain is a distributed ledger used to store information upon which a large community agrees without having a central authority that coordinates the consensus. Avoiding going into the details of how the consensus protocol works, what matters for this paper is the accounting standard that is used to store the data on the blockchain. There are several alternatives that are currently being used, but the most popular so far - and first to be implemented in Bitcoin - is the Unspent Transaction Output (UTXO). 

\begin{figure}[t]
    \centering
    \includegraphics[width=\textwidth]{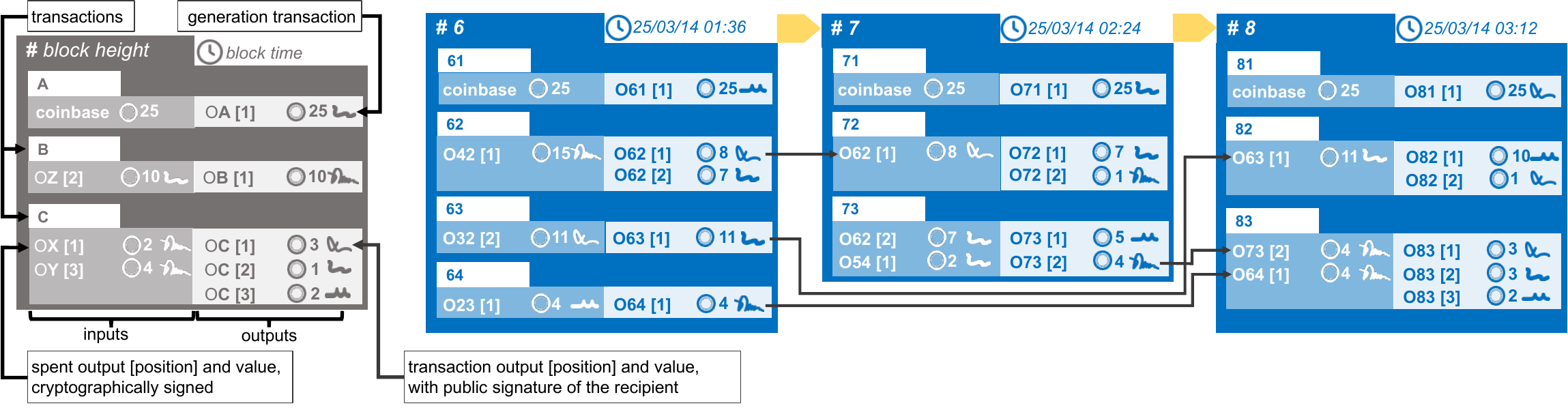}
    \caption{Schematic representation of a UTXO blockchain. Each block contains transactions, in turn consisting of a set of inputs and outputs. Total input and output values must be matching. An output grants usage rights of some amount of tokens to an address, identified by its public key. Inputs are references to unspent outputs from previous blocks, cryptographically signed by matching the public and private keys. A coinbase transaction generates new tokens as reward to the miner of the block.}
    \label{fig:utxo}
\end{figure}

In this protocol, schematised in Fig. \ref{fig:utxo}, information is stored on the blockchain in form of \textit{transactions}, transferring usage rights of some quantity of digital asset from one cryptographic \textit{address} to another. These addresses consist of two parts: a \textit{public key}, visible to everybody and used for verification, and a \textit{private key}, which is information only available to the address owner. When an agent wants to transfer some assets to another agent, they need to point to a previously generated transaction output that has not been spent yet and, using their private key, demonstrate that they own the address that output was destined to. This shows they own the usage rights on the assets in that output, and they are able to transfer them by pointing to an address belonging to the new transaction receiver.

Since all this information is stored on the blockchain, this means that the time occurring between the generation of the UTXO and its spending transaction is easily available, thus allowing to measure $\tau$ for all coins without further assumptions. These can then be used to estimate $p_i(\tau)$ depending on ownership, thus providing all the necessary information to calculate the MicroVelocity of Eq. \ref{eq:microvel}.

\paragraph*{Measuring MicroVelocity on blockchains.}
Events are recorded on the blockchain once every few minutes, when a new block is added. The consequence is that time is effectively discretised, and $\tau$ can be expressed in terms of number of blocks between transactions, $\tau \in \mathbb{N} \setminus \lbrace 0 \rbrace$. Following the logic above, let us define $P_i(\tau)$ to be the probability mass function of the discrete holding times for user $i$'s coins. Then $M_i P_i(\tau)$ coins by user $i$ have holding time $\tau$, resulting at stationarity in a contribution to user $i$'s flow of

$$
F_i(\tau) = \frac{1}{\tau} M_i P_i(\tau)
$$




It then follows that the total velocity, given in continuous time by Eq. \ref{eq:microvel}, turns into the following formula in block time

\begin{equation}\label{eq:velocity}
    V = \sum_{i=1}^N \sum_{\tau=1}^{\infty} \frac{M_i}{M} \frac{1}{\tau} P_i(\tau)
\end{equation}

Isolating the $i$-th component of this sum we are then able to define user $i$'s contribution to the macroeconomic velocity of money, i.e. the MicroVelocity, which is

\begin{equation}\label{eq:microvel_d}
    V_i = \sum_{\tau=1}^{\infty} \frac{M_i}{M} \frac{1}{\tau} P_i(\tau)
\end{equation}

The question is then how to measure $P_i(\tau)$ to be able to compute these quantities. For the specific case of UTXO-based cryptocurrencies, a very simple estimator can be adopted: the accounting protocol itself produces the holding time $\tau$ of each unit of currency as a by-product, hence the empirical probability of these holding times conditional on the agent can be estimated. For example, in Fig. \ref{fig:utxo}, the inputs of transaction \texttt{83} have the same value (4 tokens) but different holding time of two blocks (\texttt{O64[1]}) and of one block (\texttt{O73[2]}). This means, according to Eq. \ref{eq:microvel_d}, that tokens in \texttt{O64[1]} generate half as much velocity as the ones in \texttt{O73[2]}.

There are two further cautions that need to be mentioned. The first is that in principle there is no guarantee that a transaction recorded on the blockchain is actually a transfer of value from an agent to another: addresses do not have a one-to-one correspondence with agents' wallets, and in fact a wallet can contain an arbitrary number of addresses. Moreover, by protocol, inputs of transactions have to be spent in full each time: the total value of inputs has to match the total value of outputs, and the value of inputs is constrained by the UTXO they point to. The way in which this is solved is by generating an extra output, called \textit{change output}, which points the excess input to an address which is actually owned by the sender. These transactions do not actually move funds between individuals, thus they should be excluded from the calculation of MicroVelocity. 

To solve this issue, when we find an output that is likely to belong to the transaction sender, we do not count it on the MicroVelocity and add the weighted average holding time of the inputs of that transaction to its holding time. 
There are several ways to identify self-transactions through heuristic methods which take advantage of typical patterns used by cryptocurrency wallet clients. We provide further details about this in the Materials and Methods section. 

The second caution that is needed is that, when calculating MicroVelocity, the measurable holding times are right-censored, or in other words one would need future information to know the holding time of current UTXOs. While there is no straightforward solution to this, it is in principle possible to devise more refined statistical methods to infer $P_i(\tau)$ so that one can reduce this issue, while also controlling for the impact of non-stationarity in the estimation, which we have neglected in this work. We choose not to tackle this problem, as it would require some arbitrary assumptions, and defer the exploration of solutions to future research. 

\section*{Materials and Methods}

\paragraph{Data processing.}
We directly collect data from the Litecoin, Dogecoin, Monacoin and Feathercoin blockchains by synchronizing full nodes in archive mode. The complete history has then been preprocessed using the Blocksci Python library \cite{kalodner2020blocksci}, which turns raw data files into an efficiently searchable database structure.

There are many heuristic methods that have been proposed to assign addresses to unique wallets. Here we combine three heuristics, the MultiInput (MI), the Change Address (CA) and the Peeling Chains (PC):
\begin{itemize}
    \item MI takes advantage of the fact that multiple addresses appearing as inputs of the same transaction have a high likelihood of belonging to the same agent, as they had to be signed with all the private keys together. For example in Fig. \ref{fig:utxo}, \texttt{O64[1]} and \texttt{O73[2]} are assigned to the same wallet since they appear together as inputs of transaction \texttt{83};
    \item CA follows from the argument that, when there is more than one input and more than one output in a transaction, it is likely that the smallest output is a change address, provided it is smaller than the smallest input (otherwise such input would have not been necessary) and that it is a new address; if one such output is identified, it is considered to belong to the same agent as the input addresses. In Fig. \ref{fig:utxo}, output \texttt{O83[3]} is likely a change address and is associated with \texttt{O64[1]} and \texttt{O73[2]};
    \item Finally, PC identifies transaction patterns called peeling chains: sequences of transactions with a single input and two outputs, that arise when an agent needs to spend a large value UTXO (like a coinbase transaction), generating chains of transactions where the UTXO is slowly consumed. In order to identify a peeling chain, at least two subsequent transactions have to match this pattern. If that is the case, the outputs that connect the chain are assigned to the same wallet as the inputs. In Fig. \ref{fig:utxo}, transactions \texttt{62} and \texttt{72} constitute a peeling chain, hence addresses \texttt{O42[1]}, \texttt{O62[1]} and \texttt{O72[1]} are linked with one wallet.
\end{itemize}

When using heuristics to cluster together output and input addresses, there is a significant risk of merging multiple wallets together. For this reason we adopt some logical selection rules to limit the impact of this issue. In particular we apply the following conditions:

\begin{itemize}
    \item if different heuristics disagree on the change address, no change address is selected;
    \item if all heuristics agree on the change address or select none, that address is considered the change;
    \item if an input address appears also in the output, it is necessarily the change and all heuristics are neglected.
\end{itemize}

In the Supplementary Information we check the robustness of the results presented in the following sections against the choice of address clustering method.

\paragraph{Estimation of MicroVelocity.} The estimation of MicroVelocity follows from the sum of Eq. \ref{eq:microvel_d}. At a given time $t$, an agent owns $M_i(t)$ coins, each with some holding time $\tau$ that is estimated as the age of the UTXO they are included in when they are spent. Then the time $t$ probability mass function of holding times of agent $i$, $P_i^t(\tau)$, is measured as the empirical probability

$$
P_i^t(\tau) = \frac{w_i(\tau)}{M_i(t)}
$$
where $w_i(\tau)$ is the total value of UTXOs with holding time $\tau$, ensuring that $\sum_\tau w_i(\tau) = M_i(t)$. Plugging this into Eq. \ref{eq:microvel} gives the empirical value of agent $i$'s MicroVelocity at time $t$
$$
V_i(t) = \sum_\tau \frac{1}{\tau} \frac{w_i(\tau)}{M(t)}
$$
where $M(t)$ is the total amount of money in the system at the time. Notice that this value is known by construction, since the supply of cryptocurrencies is deterministic. We neglect additional considerations on the definition of circulating supply, as they would require arbitrary assumptions and we believe they would have a marginal impact on our results. Indeed, since the holding time $\tau$ appears at the denominator of Eq. \ref{eq:microvel_d}, coins with a longer holding time that may be classified in less liquid monetary aggregates already contribute less to the velocity.

\section*{Results}

\begin{figure}
    \centering
    \includegraphics[width=\textwidth]{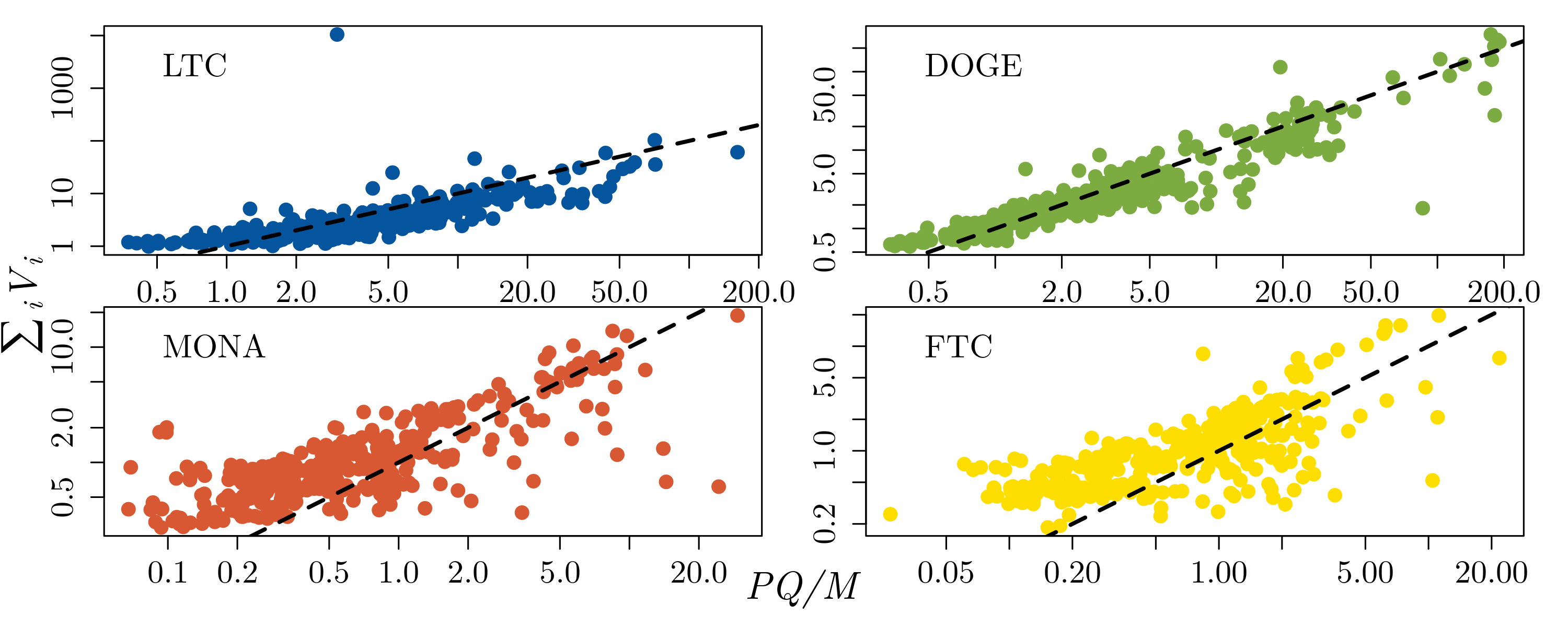}
    \caption{Comparison of total MicroVelocity and the ratio of total flow to monetary mass $PQ/M$, with weekly aggregation. All quantities are annualised.}
    \label{fig:fishervel}
\end{figure}

\paragraph{The distribution of MicroVelocity.} We begin our analysis by checking if the total MicroVelocity we measure correctly matches the velocity calculated following Fisher's Eq. \ref{eq:fisher}. In Figure \ref{fig:fishervel} we compare the sum of MicroVelocities $\sum_i V_i$, averaged weekly, with the average volume of transactions $PQ$ divided by the total money supply $M$ on that week. Both quantities are annualised to facilitate comparison. It appears clear that Eq. \ref{eq:velocity} holds true, with the total MicroVelocity even producing a slightly less volatile estimate of $V$. This is not surprising, since MicroVelocity accounts for components of $V$ at all timescales whereas $PQ/M$ only considers transactions happening on a specific time window (in this case a week), thus being more prone to temporary fluctuations.

However, as discussed above, the advantage of having this microeconomic picture is that we are able to investigate the distributional aspects of the velocity of money. We then proceed to take weekly cross-sectional snapshots of the agents' $V_i$s, which we then utilize to extract a number of statistics and test hypotheses. First, we remove values of $V_i(t)=0$ from our analysis. These may occur for multiple reasons: the agent may have not entered the market yet or has quit, thus they will have $0$ wealth and velocity, but it may also be that they own coins which they never spend. The latter can itself be caused by a behavioral choice of holding the coins indefinitely - the so-called ``HODLers" -, by the right-censoring of data we previously discussed or by technical problems such as lost passwords or hacks: this uncertainty leads us to exclude also agents with non-zero wealth that have zero MicroVelocity.

\begin{figure}[t]
\centering
\includegraphics[width=.35\textwidth]{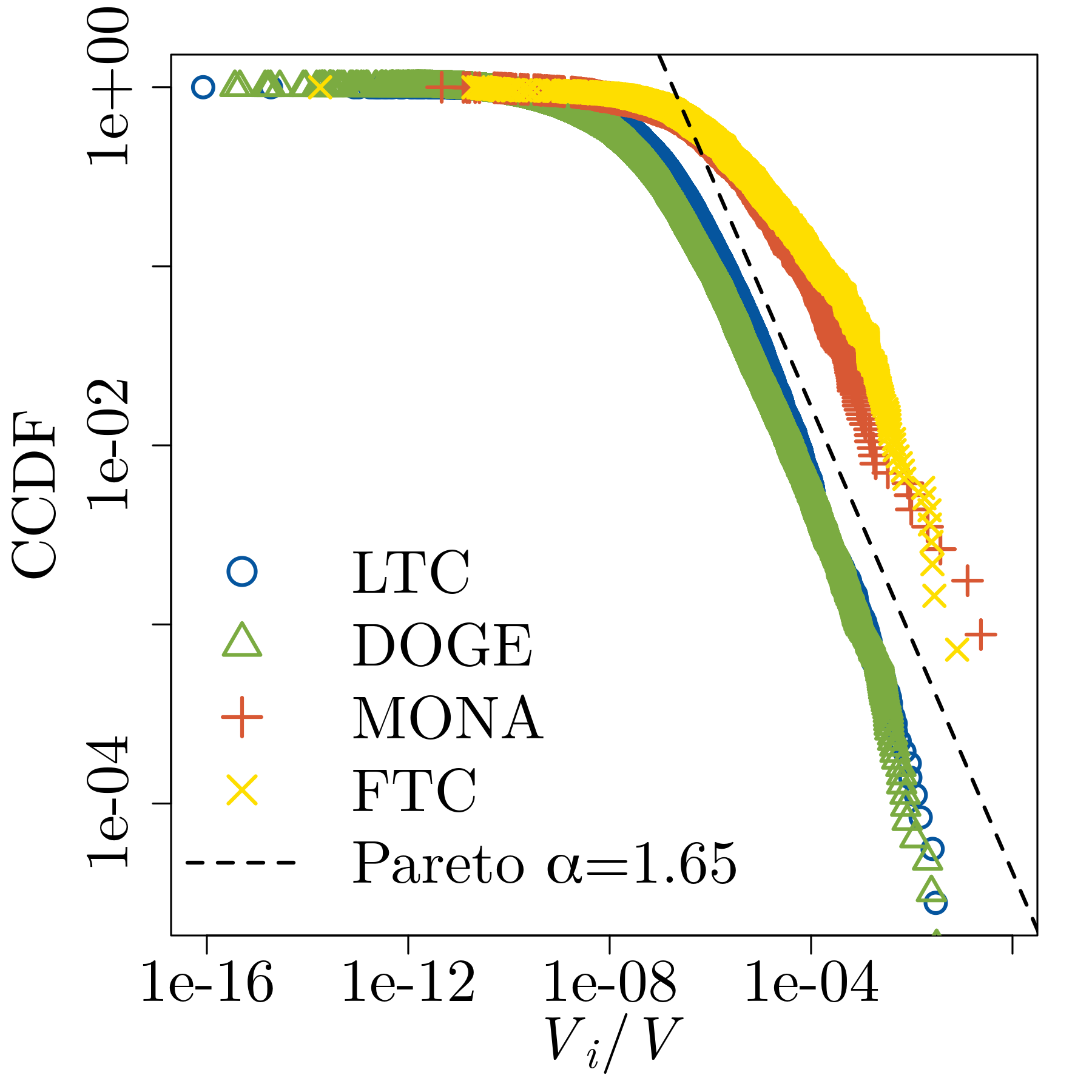}
\resizebox{0.64\textwidth}{!}{\begin{tabular}[b]{r|rrrrr}
\hline
& & LTC & DOGE & MONA & FTC \\ 
  \hline
\multirow{6}{*}{$\alpha$} & Mean & 1.69 & 1.65 & 1.56 & 1.62 \\ 
 & Min. & 1.44 & 1.52 & 1.28 & 1.27 \\ 
 & Q1 & 1.64 & 1.61 & 1.52 & 1.52 \\ 
 & Median & 1.70 & 1.65 & 1.55 & 1.59 \\ 
 & Q3 & 1.74 & 1.68 & 1.61 & 1.69 \\ 
 & Max. & 1.92 & 1.80 & 2.29 & 2.20 \\ 
   \hline
   \hline
 \multirow{2}{*}{$D_{Par}$} & Mean & 0.03 & 0.03 & 0.08 & 0.11 \\ 
 & $p > 0.2$ & 92.96\%  & 100.00\%  & 99.92\%  & 94.07\%  \\ \hline
 \multirow{2}{*}{$D_{Exp}$} & Mean & 0.83 & 0.86 & 0.83 & 0.78 \\ 
 & $p > 0.2$ & 0\% & 0\% & 0\% & 0\% \\ 
   \hline
\end{tabular}
}
\captionlistentry[table]{}
    \captionsetup{labelformat=andtable}
    \caption{(left) Snapshot of complementary cumulative distribution function (CCDF) of relative MicroVelocity $V_i/V$ on April 4th, 2016 for the four cryptocurrencies. A Pareto distribution with tail exponent $\alpha=1.65$ is reported as comparison. (right) Pareto fit and Kolmogorov-Smirnov test. Summary statistics - mean, quartiles, minimum and maximum - for the estimated $\alpha$ parameter of the Pareto distribution and for the K-S tests. The ``$p>0.2$" lines report the percentage of two-tailed K-S tests that don't reject the Pareto and Exponential nulls at the $20\%$ confidence level. Critical values obtained by bootstrap.}\label{tab:KS}
\end{figure}

We then proceed to analyze the cross-sectional distribution by means of the Kolmogorov-Smirnov distance $D$ from a known distribution. This is defined as
$$
D = \sup \lvert \hat{P}(V_i) - P(V_i) \rvert
$$
where $\hat{P}(V_i)$ is the empirical distribution of MicroVelocities and $P(V_i)$ is an arbitrary probability distribution. This distance is the test statistic for the two-tailed Kolmogorov-Smirnov test, which tests the null hypothesis that $\hat{P}(V_i) \equiv P(V_i)$. Since $V_i \geq 0$ by definition, we choose to take as comparisons the exponential and Pareto distributions, to gain insight on whether the data is closer to a thin- or to a fat-tailed distribution. We fit the parameters of these distributions by Maximum Likelihood methods and calculate the distances $D_{Par}$ and $D_{Exp}$, obtaining the results summarized in Table \ref{tab:KS}. While the thin-tailed exponential null hypothesis is always rejected by the Kolmogorov-Smirnov test, it is clear that a fat-tailed Pareto distribution with tail exponent $\alpha \approx 1.6$ is a much better descriptor for the distribution generating the data. Notice that we choose a very high significance threshold of $0.2$, since we want to check the robustness of the null hypothesis rather than identifying an alternative. Despite this, the test almost never rejects the Pareto null, consolidating the observation that the total velocity of money is very unevenly distributed across agents, with a small minority dominating the sum of Eq. \ref{eq:microvel_d}. 

\begin{figure}[t]
    \centering
    \includegraphics[width=\textwidth]{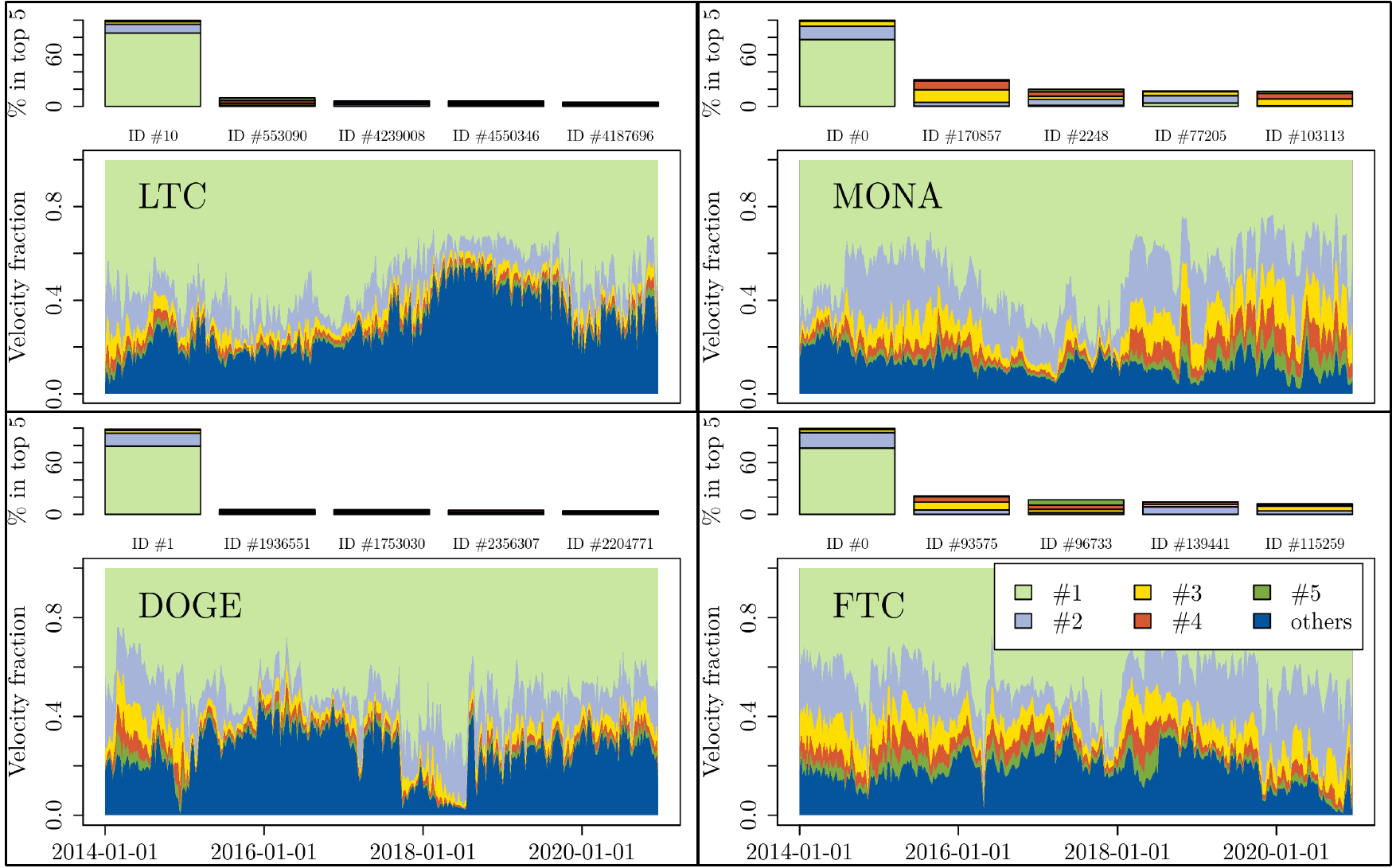}
    \caption{Area charts of MicroVelocity and identity of top 5 individual contributors for the four cryptocurrencies, averaged weekly. In each panel the top chart shows the frequency (vertical axis) with which the 5 most represented agents (horizontal axis) figure in the top 5 contributors, with stacked bars colored to show their ranking according to the legend. The bottom charts show the share of $V$ contributed by the top 5 agents as a function of time. We see that one agent is always in the top 5 and that in all cases the vast majority of the total velocity is concentrated in the hands of a few agents.}
    \label{fig:inequality}
\end{figure}

Perhaps the most striking evidence of this comes from the stacked charts of Fig. \ref{fig:inequality}. In each panel's bottom chart, each colored area corresponds to the fraction of total velocity that is produced by the 5 largest contributors, compared with the total contributions by all others, tracked over time with weekly aggregation. In all the analyzed cryptocurrencies the vast majority of the total velocity is generated by less than 5 agents: most likely these - whose actual identity is unknown to us - are companies that act as intermediaries, such as exchange markets, custodians or layer 2 payment services like the Lightning Network \cite{lin2020lightning}. Despite the real-world identity of these agents being unknown, we can still track them by their ID from the address clustering algorithm, which means we are able to count how frequently each agent features as a top contributor. The top charts of Fig. \ref{fig:inequality} show exactly this, counting the percentage of weeks a given agent is among the top 5 and showing their ranking with the colored bar. It appears that only one agent is always in the top 5, most of the times as top contributor, whereas others show significantly less persistence. This is most likely due to our choice of heuristics for address clustering, which possibly clusters together multiple exchanges.

\begin{figure}[t]
    \centering
    \includegraphics[width=\textwidth]{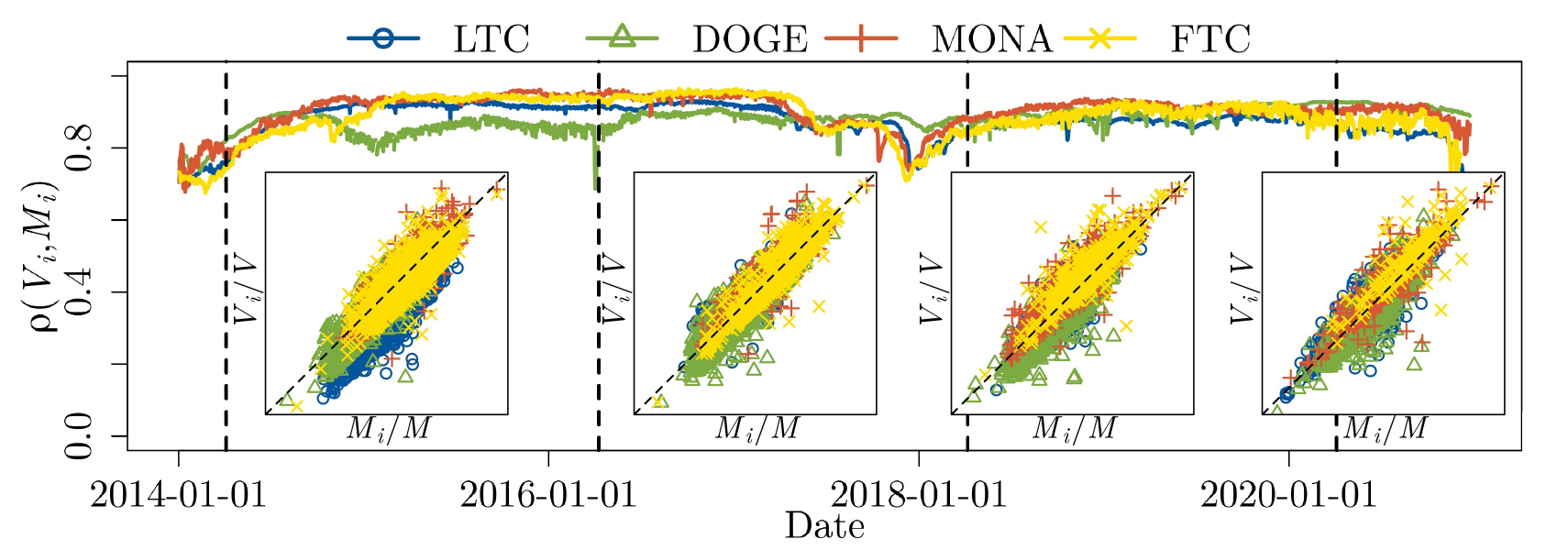}
    \caption{Spearman's rank correlation between the MicroVelocity $V_i$ and the wealth $M_i$ of the agent. Insets: scatterplots of ${V_i}/{V}$ vs ${M_i}/{M}$ in log-log scales at marked dates, with dashed line on the diagonal ${V_i}/{V} = {M_i}/{M}$.}
    \label{fig:corr}
\end{figure}

\paragraph{MicroVelocity and wealth.}
While we have no information about the real-world identities of agents, we can still characterize them by their activity on the blockchain and see how their features correlate with their MicroVelocity. In particular we focus on the relation arising between MicroVelocity and agent wealth $M_i$. As mentioned when we introduced Eq. \ref{eq:microvel}, in an economy where the ``consumption functions" incorporated $P_i(\tau)$ are homogeneous, $M_i$ would be the only factor to generate diversity in MicroVelocity. The systems we analyze have high wealth concentration, which is a tendency that is common to cryptocurrencies \cite{kondor2014rich, hileman2017global,vallarano2020,decollibus2021}. This is likely due to large pseudo-banking companies that arise naturally in the environment, acting as exchange markets for fiat currencies or as vaults where crypto users securely store their tokens \cite{sai2021taxonomy}. In Fig. \ref{fig:corr} we show the Spearman's rank correlation coefficient $\rho$ between $V_i$ and $M_i$ over time. We choose Spearman's coefficient to overcome the limitation of Pearson's correlations for highly heterogeneous data (as discussed in the previous section), where outliers would taint the estimation of the correlation. Fig. \ref{fig:corr} shows that the correlation is positive and consistently above 0.6. While it is clear that $M_i$ is a strong determinant of $V_i$, as expected from Eq. \ref{eq:microvel}, there is still a significant variance around the linear trend, as we show in the inset scatterplots, and also variation over time. This residual variance is then entirely due to the heterogeneity in $P_i(\tau)$, which does not seem to be itself dependent on wealth. One interesting phenomenon seems to appear during late 2017, with $\rho$ decreasing throughout the period of build-up of the crypto bubble that peaked in December 2017 \cite{kyriazis2020systematic}.

\paragraph{MicroVelocity and the structure of the economy.}
We conclude our analysis by considering the role of MicroVelocity in the structure of transaction patterns, introducing transaction networks. A network is an ordered pair of sets $G=(\mathcal{V},\mathcal{E})$, called the nodes $\mathcal{V}$ and edges (or links) $\mathcal{E}$, where the elements of $\mathcal{E} \in \mathcal{V} \times \mathcal{V}$ represent connections between elements of $\mathcal{V}$. Identifying agents as nodes and transactions as directed connections from the transaction sender to the receiver, as done for instance in \cite{bovet2019evolving}, it is possible to obtain a topological description of the economy, highlighting with whom agents exchange tokens and how central they are in the flow of value in the economy. We construct transaction networks on weekly aggregation, which is the minimum timescale that allows to mitigate the effect of day-of-the-week seasonalities that are common in economic and financial data (e.g. weekend effects, higher volumes on Mondays/Fridays, etc.). Operationally this means that a connection from agent $i$ to agent $j$ is present if at least one transaction occurred between them in a given week. We then proceed to take the values of $V_i$ at the end of each week and assign them to each node as attributes, which then allows us to consider the assortativity coefficient of these networks with respect to MicroVelocity.

\begin{figure}[t]
    \centering
    \includegraphics[width=\textwidth]{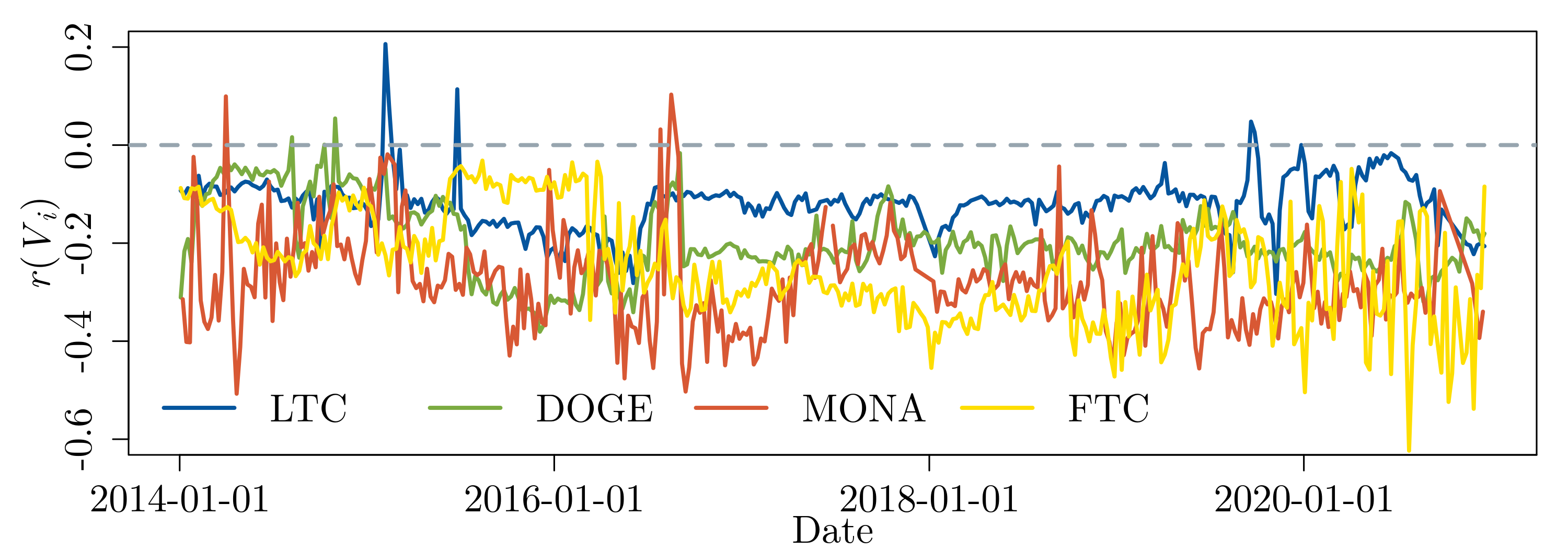}
    \caption{Rank assortativity of MicroVelocity on weekly transaction networks.}
    \label{fig:assort}
\end{figure}

The assortativity coefficient, defined by \cite{newman2002assortative}, is the Pearson correlation coefficient of a given node-specific quantity across linked pairs in a network. It has been introduced to measure how nodes select their neighbors with respect to specific characteristics, most prominently the degree, and it is defined as
$$
r = \frac{\sum_{(i,j) \in \mathcal{E}} ( x(i) - \langle x \rangle) (x(j) - \langle x \rangle )}{\sum_{i \in \mathcal{V}} (x(i) - \langle x \rangle)^2}
$$
where $x(i)$ is the node-specific property of interest and $\langle \cdot \rangle$ indicates the arithmetic mean operator. A positive value of this coefficient indicates homophily, i.e. links mainly connect similar nodes in terms of the selected property, whereas a negative value indicates heterophily, i.e. connections mostly appear between nodes with very different values of $x$.

In our case we decide to use the rank in MicroVelocity as the nodes attribute $x$, thus computing a Spearman's correlation instead of Pearson's, again to account for the fat-tailed nature of the distribution of $V_i$s. We report our finding in Fig. \ref{fig:assort}, where we plot this rank assortativity coefficient for MicroVelocity on weekly networks as a function of time. The resulting measure is mostly negative, thus suggesting that highly-ranked nodes mostly trade with lowly-ranked ones and vice-versa. 

\section*{Discussion}

In this article we investigated the microeconomic foundations of the macroeconomic velocity of money, building upon theoretical work \cite{wang2003circulation} and realizing the first empirical measurement of the velocity of money from micro-level data. We leveraged on the technological breakthrough provided by digital currencies and by cryptocurrencies in particular, which offer unprecedented detail about the movement of money in a closed economy where all coins can be followed in every transaction they take part in. We have thus been able to introduce the individual contribution to velocity by single agents, which we have called MicroVelocity, as a function of the agent-conditional distribution of holding times of money. 

We argue that insights about the distribution and characterization of MicroVelocity are valuable to researchers, policymakers and to the public to gain a deeper understanding of the structure of the economy: in the particular case at hand, we provide evidence that cryptocurrency economies' claim of being ``decentralized" is largely unjustified, as shown by the extremely skewed distributions of wealth and MicroVelocity we measure. This ``decentalization illusion" \cite{aramonte2021defi} is unmasked by the highly heterogeneous MicroVelocity distribution, indicating that few entities are intermediaries to most of the transactions, since large values of MicroVelocity are easily due to high turnover in assets. 

The fact that we find an extremely positive correlation between MicroVelocity and wealth, as well as negative assortativity of $V_i$ in transaction networks, is also corroborating evidence that these economies have seen the rise of pseudo-banking services which centralize the supply of money and act as custodians and intermediation services, often without any scrutiny by regulators due to the lack of appropriate legislation. 

We envision multiple directions that could be taken following this work, tackling some limitations of our study as well as expanding its applicability beyond the relatively narrow realm of UTXO-based cryptocurrencies. 

The specific implementation we proposed here takes advantage of the fact that holding times are relatively easy to compute thanks to the age of UTXOs, but this can be easily overcome in the case where this information is not available by considering, for instance, a Last-In-First-Out (LIFO) or First-In-First-Out (FIFO) spending rule, as done for instance in \cite{mattsson2019networks}. We argue that a LIFO rule is possibly the most economically significant for a medium of exchange, as it would automatically exclude less liquid portions of the money supply from the calculation, but we leave this extension for future research.

Our approach using empirical probabilities to estimate $P_i(\tau)$ has a limitation in the fact that holding times data is right-censored, and this becomes a bigger issue the closer one is to present times. Moreover using empirical probabilities does not fully respect the assumptions on stationarity; nonetheless the convergence between the micro and macro quantities shown in Fig. \ref{fig:fishervel} leads us to think that they still are a good first approximation. We are confident that these issues could be reduced by performing some additional assumptions on the form of the holding times distribution, possibly giving it a parametric form justified by theory (e.g. the Gamma family proposed in \cite{wang2003circulation}). This would mitigate the limitations and possibly allow to produce forecasts as well as predict the outcome of policy interventions.

\bibliographystyle{unsrt}
\bibliography{biblio.bib}

\end{document}